\documentstyle[prl,aps,multicol]{revtex}
\renewcommand{\narrowtext}
{\begin{multicols}{2} \global\columnwidth20.5pc}
\renewcommand{\widetext}
{\end{multicols} \global\columnwidth42.5pc}

\begin{document}

\newcommand{\be}{\begin{equation}}
\newcommand{\ee}{\end{equation}}
\newcommand{\bea}{\begin{eqnarray}}
\newcommand{\eea}{\end{eqnarray}}
\newcommand{\nt}{\narrowtext}
\newcommand{\wt}{\widetext}

\title{Entanglement and decoherence in near-critical qubit chains}

\author{D. V. Khveshchenko}

\address{Department of Physics and Astronomy, University of North
Carolina, Chapel Hill, NC 27599}
\maketitle

\begin{abstract}
We study the problem of environmentally-induced decoherence in a near-critical
one-dimensional system of $N\gg 1$ coupled qubits. Using
the Jordan-Wigner fermion representation of the qubit operators we identify
the decoherence rates relevant for the two-qubit reduced density matrix. 
We find that a desirable onset of massive shared entanglement 
in the near-critical regime comes at the expense of decoherence 
which also tends to increase as the system is tuned towards criticality.
Our results reveal rather contradictory general requirements
that future designs of a qubit chain-based quantum information processor will need to satisfy.
\end{abstract}

\nt 
Besides establishing a number of other, previously unexplored,
connections, the rapidly developing field of quantum computing
brings about new links between statistical mechanics of quantum
spin systems and quantum information theory.

This relationship was originally prompted by the idea of using
electron or nuclear spins as natural candidates for individual
physical qubits and constructing a practical quantum register for
implementing various quantum protocols.
Moreover, it is believed to have an even greater potential, as far
as practical control of quantum entanglement and coherence is concerned. 

To this end, a
number of authors have put forward the idea of constructing novel types of logical qubits 
which can enjoy an exceptionally high degree of coherence, thanks to the robust intrinsic
correlations present in quantum states of many interacting spins.
Instead of considering the qubit interactions
as a nuisance to be rid of, these proposals seek to utilize them, focusing
on topological and, therefore, intrinsically coherent 
Majorana fermion (edge) or anyon-like (bulk) excitations 
of some strongly correlated spin states proposed as 
a platform for implementing
fault-tolerant computations with automatically built-in error correction codes.

However, while offering a potentially ideal protection against
environmentally induced decoherence,
these proposals require a macroscopically large 
number of interacting physical qubits in order
to construct only a few (topo)logical ones, and they 
may also face general difficulties with efficient information encoding and readout.  

Alternative, more memory-efficient,
approaches to controlling decoherence, such as 
creation of decoherence-free subspaces (DFS) \cite{Zanardi}
by means of dynamical decoupling ("bang-bang" pulse control) 
\cite{Viola}, often invoke certain symmetries of the system-to-environment 
couplings and/or rely on the
possibility of switching the qubits' interactions on and off at will,
although, in practice, either condition may not necessarily be easy to meet.

Regardless of the actual physical makeup of the qubits,
most of the the previous analyses have been based on the generic spin-$1/2$ Hamiltonian
\be
H=\sum_{a=x,y,z}{\ (}\sum_{i=1}^N{B}^a_i(t){S}^a_i+\sum_{i,j=1}^NJ^a_{ij}(t){S}^a_i{S}^a_j{\ )}
\ee
and an abstract quantum computing protocol has been thought of as a sequence of 
short pulses implementing one-qubit $B^a_i(t)=B_i^a\theta(t)\theta(\tau-t)$, and two-qubit,
$J^a_{ij}(t)=J_{ij}^a\theta(t)\theta(\tau-t)$, gate operations.

It has been recognized, however, that any realistic qubits would interact with each other
not only during the externally controlled gate operations but also
during the idling periods due to their unwanted 
short-ranged (such as exchange, $J^a_{ij}\propto\exp(-const|i-j|)$)
or long-ranged (such as dipolar, $J^a_{ij}\propto 1/|i-j|^3$) static couplings.

In this Letter, we study the possibility
of utilizing these residual interactions for creating and maintaining
quantum entanglement between the individual qubits that are also exposed
to a noisy environment ("bath").

As one analytically tractable, yet sufficiently non-trivial,
example we consider a system of $N\gg 1$ qubits with the topology of a 
one-dimensional $(1D)$ chain and time-independent 
nearest-neighbor couplings, $J^a_{ij}=-J^a\delta_{j,i\pm 1}$,
while describing the bath by a fluctuating component of the 
local magnetic field, ${B}^a_i(t)={B}^a+{h}^a_i(t)$.

Specifically, we focus on the two-parameter set of the Hamiltonians with
$B^z=B, ~~J^{x,y}=J(1\pm\gamma),~~B^{x,y}=J^z=0$ 
which is exactly solvable via the
Jordan-Wigner (JW) transformation from spin-$1/2$ operators to spinless
1D fermions 
\be 
S_i^z=\chi^\dagger_i\chi_i-1/2, ~~~
S_i^{-}=S_i^{x}-iS^{y}_i=\chi_i\prod_{j=1}^{i-1}e^{i\pi \chi^\dagger_j\chi_j}
\ee
Expressed in terms of the JW fermions, Eq.(1) takes on the form
\be
H={1\over 2}\sum_{i=1}^N{\ (}B\chi^\dagger_i\chi_{i}-J
\chi^\dagger_i\chi_{i+1}-J\gamma \chi_i\chi_{i+1}{\ )}+
h. c. 
\ee
that can be diagonalized by virtue of the Fourier transformation
$\chi_j=\sum_{k}e^{ikaj}\chi_k$ (here $k=\pm 2\pi n/aN$
is the JW fermion momentum
inversely proportional to the lattice spacing $a$), followed by the subsequent 
Bogoliubov rotation
$\chi_{\pm k}=\cos\theta_k\psi_{\pm,k}\pm\sin\theta_k\psi^\dagger_{\mp,k}$
through the angle defined by the relation
$\tan 2\theta_k=J\gamma\sin ka/(B-J\cos ka)$ which turns (3) into a sum 
$H=\sum_{\pm, k}\epsilon_{k}\psi^\dagger_{\pm, k}\psi_{\pm, k}$
over single-fermion eigenstates $\psi_{+, k}$
and their charge conjugates ("holes") $\psi^\dagger_{-, k}$ with the dispersion 
\be
\epsilon_k={\sqrt {J^2\gamma^2\sin^2ka+(B-J\cos ka)^2}}
\ee
At $0<\gamma\leq 1$ the Hamiltonian (3) belongs to the universality class
of the Ising model in transverse field, 
featuring a pseudo-relativistic low-energy dispersion 
$\epsilon_k\approx{\sqrt {{v}^2k^2+\Delta^2}}$.
The spectral gap (JW fermion "mass") $\Delta=|B-J|$ vanishes
at the critical point $\lambda=B/J=1$,
whereas the velocity $v=J\gamma a$ remains finite.
By contrast, at $\gamma=0$ corresponding to 
the $XY$-model the fermion dispersion is quadratic at small momenta.

In the thermodynamic ($N\to\infty$) limit, the model (3) demonstrates 
an onset of massive entanglement upon approaching its critical point \cite{Amico}, 
as quantified by the concurrence ${\cal C}_{ij}=max[0, r_1-r_2-r_3-r_4]$
defined in terms of the (numbered in the descending order) square roots 
$r_\alpha$ of the eigenvalues  of the product between the two-qubit 
reduced density matrix (RDM) ${\hat \rho_{ij}}=Tr_{{\overline i},{\overline j}}{\hat \rho}$
and its time-reversal conjugate
${\hat \sigma}^y_i\otimes{\hat \sigma}^y_j{\hat \rho}^*_{ij}{\hat \sigma}^y_i\otimes
{\hat \sigma}^y_j$ \cite{Wooters}.

Near the critical point the nearest- and next-nearest-neighbor
concurrences increase and their derivatives
exhibit logarithmic singularities ${d{\cal C}_{i,i\pm 1}/d\lambda}
\propto {d^2{\cal C}_{i,i\pm 2}/d\lambda^2}
\propto\ln |\lambda-\lambda_c|$ with $\lambda_c\approx 1$ \cite{Amico},  
consistent with the effective description in terms of
the free (pseudo)relativistic $1D$ fermions with mass $\Delta\propto |\lambda-\lambda_c|$.

The tendency of the correlated spin-liquid ("resonant valence bond") 
states to become more and more entangled in the critical
(gapless) regime has to be distinguished from the
robust singlet correlations characterizing 
gapful "valence bond solid" states where, instead of being shared by many
qubits, the entanglement is confined to their designated pairs.

In light of the fact that distrubuted entanglement is considered an important 
resource for quantum information processing, one may be inclined to conclude that 
in order to take a full advantage of the increasing qubit correlations 
the system has to be tuned into the critical (gapless) regime, e.g., by
adjusting the uniform field ($B\to J$).

However, any discussion of the potential benefits of the desired massive
entanglement between the qubits would be incomplete without  
assessing the possibility of controlling their 
(in contrast to the former, unwanted) entanglement with the
bath which constitutes a source of environmentally-induced decoherence.
In this regard, one of the key questions is whether or not this can be done 
by virtue of the same inter-qubit couplings
that have facilitated the onset of the critical entanglement in the first place. 

In what follows, we choose the qubit-bath coupling Hamiltonian in the 
form $X=\sum_{i=1}^NS^z_ih^z_i(t)$. 
Unlike in the case of non-interacting qubits, the latter causes not only pure dephasing 
(phase errors) but also relaxation (bit-flip errors), since $[H,X]\neq 0$.
A spatial inhomogeneity of a generic $D$-dimensional
dissipative environment comprised of bosonic modes
(photons, phonons, spin waves, etc.) with a dispersion $\Omega_q\propto q^\beta$
requires one to use a dynamic spectral function 
$$
D^{ab}(\omega, {\vec q})=\sum_{i,j=1}^NIm\int^\infty_0 dt 
e^{i\omega t-i{\vec q}{\vec x}_{ij}}<h^a_i(t)h^b_j(0)>
$$
\be
\propto \delta^{ab}\omega^{\alpha+2-D}\delta(\omega^2-\Omega_q^2)
\ee
where the vector ${\vec x}_{ij}$ of length $a|i-j|$ is parallel to the chain, instead of a
total spectral density $\int D^{zz}(\omega, {\vec q})d^D{\vec q}$. 

Motivated by the results of Refs.\cite{Amico} pertinent to the pairwise
entanglement and considering the special importance of two-qubit gates in all the previously
proposed quantum protocols, we will be primarily concerned with the two-qubit RDM which  
can be expanded over the one- and two-spin correlation functions
$$
{\hat \rho}_{ij}(t)={\bf 1}_i\otimes{\bf 1}_j
+\sum_{a}(<S^a_i(t)>{\bf 1}_i\otimes{\hat \sigma}^a_{j}+
$$
\be
<S^a_j(t)>{\hat \sigma}^a_{i}\otimes{\bf 1}_j)+
\sum_{a,b}<S^a_i(t)S^b_j(0)>
{\hat \sigma}^a_{i}\otimes{\hat \sigma}^b_{j}
\ee
In particular, we will be interested in the off-diagonal RDM element 
\bea
<\downarrow\uparrow|{\hat \rho}_{ij}(t)|\uparrow\downarrow>=<S^{+}_i(t)S^{-}_j(0)>=\nonumber\\ 
<\chi_i(t)\prod_{k=1}^{i-1}e^{-i\pi\chi^\dagger_m(t)\chi_m(t)}
\prod_{l=1}^{j-1}e^{i\pi\chi^\dagger_n(0)\chi_n(0)}\chi^\dagger_j(0)>
\eea
which, alongside the rest of the RDM, becomes translationally-invariant,
$<\downarrow\uparrow|{\hat \rho}_{ij}(t)|\uparrow\downarrow>=
{\rho}_{\downarrow\uparrow;\uparrow\downarrow}(i-j,t)$,
at qubit separations $1\ll |i-j|\ll N$.

Despite the fact that the dynamical correlator (7) 
appears to be rather difficult to compute even in the noiseless case
\cite{Barouch}, it gets affected by the noise in much the same way 
as the JW fermion propagator
 $<\chi_i(t)\chi^\dagger_j(0)>$,
for the effect of the stochastic field $h^z(x,t)$ on the "charged"
JW fermion operators $\chi_i$ and $\chi^\dagger_j$ dominates over that on their    
"neutral" products $\chi_m^\dagger\chi_m$. 

In turn, the amplitude (7) can be expressed in terms of the (retarded) matrix-valued
Green function ${\hat G}^R(k,t)=\theta(t)<\Psi(k,t)\Psi^\dagger(k,0)>$ 
expandable over the Pauli matrices ${\hat \tau}_i$ acting in the Nambu (particle-hole) space
of spinors $\Psi=(\psi_+,\psi^\dagger_-)$.
In the presence of a random field $h^z(x,t)$
and in the continuum limit, this Green function obeys the equation
\be
[i\partial_t-{\hat \tau}_3\epsilon_k+
{\hat \eta}_kh^z(x,t)]{\hat G}^R(k,t|h(x,t))=\delta(t)
\ee
where we introduced the matrix ${\hat \eta}_k={\hat \tau}_3\eta_{+,k}+{\hat \tau}_1\eta_{-,k}$ 
with the coefficients $\eta_{+,k}=\cos 2\theta_k$ and $\eta_{-,k}=\sin 2\theta_k$.

For a given configuration of the random field, an approximate formal  
solution to Eq.(8) can be cast in the form (see Ref.\cite{DVK} for the technical details
of this technique)
\be
{\hat G}^R(k,t|h(x,t))=\int {d\nu\over 2\pi}\int^\infty_0 ds
e^{is(\nu-\epsilon_k{\hat \tau}_3)-i\nu t}
\ee
$$
\exp[i\int{d\omega d^D{\vec q}\over (2\pi)^{D+1}}
{\hat \eta}_kh^z(\omega,{\vec q})
\int^s_0ds^\prime
e^{is^\prime({\hat \tau}_3\epsilon_{k+q}+\nu-\omega)+i{\vec q}{\vec x}}]
$$
The Fourier transform ${\hat G}^R(x,t|h)$ of Eq.(9) represents the 
time evolution operator which incorporates the effect of "bremsstrahlung" 
(multiple emission/absorption 
of soft bosonic modes to/from the bath) and the associated recoil of the JW fermions. 

With the operator of time evolution at hand, 
one can now construct the JW fermion density matrix 
$\rho_{JW}(x-y, t|h)=\int dzdw{G}_{\pm\pm}^R(x-z, t|h){\rho}_{JW}^{th}
(z-w){G}_{\pm\pm}^A(w-y, -t|h)$
where the thermodynamic
density matrix 
${\rho}_{JW}^{th}(x)=\int e^{ikx}dk[1-\tanh(\epsilon_k/2T)]/(4\pi)$ 
describes the system of free JW fermions in equilibrium with the bath at temperature $T$. 

Upon performing a Gaussian statistical average of Eq.(9) over $h^z(x,t)$, 
one arrives at the following expression for the JW fermion density matrix
of a noisy near-critical qubit chain of length $L=Na$ 
$$
{\rho}_{JW}(x, t)=\int^{L}_0dy{\rho}_{JW}^{th}(y)
\int{dk\over 2\pi}e^{ik(x-y)}\cos^2\epsilon_kt
$$
\be
\exp[-{1\over 2}\sum_{\pm}\int {d^D{\vec q}d\omega\over (2\pi)^{D+1}} 
{1-\cos(\omega-\epsilon_k\pm\epsilon_{k+q})t\over (\omega-\epsilon_k\pm\epsilon_{k+q})^2}
\ee
$$
\eta_{\pm,k}^2D^{zz}(\omega, {\vec q})
(\coth{\omega\over 2T}-\tanh{\omega-\epsilon_k\over 2T})
(1-\cos {\vec q}({\vec x}-{\vec y}))]
$$
In the absence of the qubit-bath coupling, the kernel
of the $y$-integration in Eq.(10) reads 
\be
U(z, t)={\delta(z)\over 2}-{\Delta\over 4\pi}Im\sum_{\pm}{\sqrt {z\pm 2vt\over z\mp 2vt}}
K_1(\Delta{\sqrt {z^2-4v^2t^2}})
\ee
where $z=x-y+i0$ and $K_1(w)$ is the Macdonald function of the $1^{st}$ kind.
At $\Delta=0$ and $max[z, vt]\lesssim L$ it further reduces to the expression 
$U(z, t)=[2\delta(z)+\sum_{\pm}\delta(z\pm 2vt)]/4$ which describes 
"ballistic" spreading of the initial entanglement 
at the speed $v$ away from the $i^{th}$ and $j^{th}$ qubits in both directions.

We note, in passing, that, owing to its unitary nature,
 a spatial/temporal decay of the RDM of any
exactly solvable noiseless qubit chain,
such as Eq.(3) or the ferromagnetic Heisenberg chain studied in Ref.\cite{Eberle},
is not to be interpreted as some kind of a
"sub-exponential" (especially, temperature-independent) decoherence.

In the non-interacting ($J=0$) limit 
it is instructive to compare Eq.(10) to the well-known 
exact solution for the entire $N$-qubit density matrix \cite{Reina}
\be
{\hat \rho}^{(0)}(t)={\hat \rho}(0)
\exp[-{1\over 2}\int{d^D{\vec q}d\omega\over (2\pi)^{D+1}} 
{{1-\cos\omega t}\over \omega^2}
\ee
$$
D^{zz}(\omega, {\vec q})\coth{\omega\over 2T}
\sum_{i,j=1}^N({\hat \sigma}^z_i-{\overline{\hat \sigma}}^z_i)
({\hat \sigma}^z_j-{\overline{\hat \sigma}}^z_j)
e^{i{\vec q}{\vec x}_{ij}} ]
$$
where ${\hat \sigma}^z_i$ and ${\overline{\hat \sigma}}^z_i$ act upon the
input ($t=0$) density matrix ${\hat \rho}(0)$ on the left and from the right, respectively.

By tracing out in Eq.(12) all but the $i^{th}$ and $j^{th}$
qubits, one readily obtains the formula for ${\rho}_{\downarrow\uparrow;\uparrow\downarrow}(i-j,t)$
which almost (see below) exactly coincides with Eq.(10) for $v=\theta_k=0$, thereby 
providing further support for linking the decoherence properties of the two-qubit RDM
to those of the JW fermion density matrix $\rho_{JW}(x, t)$.

In contrast to
Eq.(10), however, in Eq.(12) the Fermi distribution function for the JW fermions is missing, consistent 
with the fact that in the non-interacting limit the qubit system lacks a well-defined temperature. 
As a result, the frequency integral in Eq.(12) receives contributions from the bosonic modes
with energies $\omega>T$ that give rise to the notorious
$T=0$ decoherence, the latter being a ubiquitous feature of  
any Caldeira-Leggett-type (single-qubit) model which
is oblivious to the Pauli exclusion principle.

As follows from Eq.(12), under the conditions of "collective decoherence" 
(i.e., when the bath coherence length is much greater than $L$, 
hence $e^{{\vec q}{\vec x}_{ij}}\approx 1$) 
the matrix element ${\rho}^{(0)}_{\downarrow\uparrow;\uparrow\downarrow}(i-j,t)$ 
demonstrates no decay at all, in agreement with the fact that 
the Hilbert space of the non-interacting qubits
possesses a DFS spanned by the states that nullify the operator  
$\sum_{i=1}^N({\hat \sigma}^z_i-{\overline{\hat \sigma}}^z_i)$ \cite{Reina}. 

In fact, the validity of the above conclusion hinges on the possibility of
characterising the bath solely in terms of the momentum-integrated spectral density.
In contrast to the case of non-interacting qubits, however, 
any non-flat JW fermion dispersion ($\epsilon_k\neq const$) 
results in a non-trivial $\vec q$-dependence of the integrand in Eq.(10), 
thus making it difficult to positively establish the existence of any DFS
even under the 
most friendly collective decoherence conditions.

Then one can readily see that for a generic encoding
it is the slowest-decaying off-diagonal matrix element 
${\rho}_{\downarrow\uparrow;\uparrow\downarrow}(i-j,t)$
that gets to control the long-time behavior of the concurrence
${\cal C}_{ij}(t)$ and other entanglement quantifiers 
such as purity $P_{ij}(t)={{Tr[{\hat \rho}^2_{ij}(t)]}}$
or fidelity $F_{ij}(t)={{Tr[{\hat \rho}_{ij}(t)e^{-iHt}{\hat \rho}_{ij}(0)e^{iHt}]}}$.

Assuming that Eq.(10) retains its basic structure 
\be
{\rho}_{JW}(x,t)=\int^{L}_0dy
U(x-y,t)\rho_{JW}(y,0)e^{-\Gamma(x-y,t)}
\ee
for an arbitrary (not necessarily thermodynamic) input density matrix 
$\rho_{JW}(x, 0)$, we now consider a practically important example of the 
isotropic $D$-dimensional bath composed of acoustic ($\beta=1$) bosonic 
modes propagating
at a speed $c\gg v$ (albeit resulting 
in no significant loss of generality, the latter condition does simplify the following analysis).

For an input state that consists of closely spaced entangled qubit pairs
($\rho_{JW}(x,0)\to 0$ for $x\gg a$),
Eq.(13) allows one to read off the proper 
decoherence rate directly from the exponential attenuation factor ("influence functional")
\bea
\Gamma(r,t)\propto\int^\infty_{0}d\omega\omega^{\alpha}\sum_{\pm}
{{1-\cos(\omega-\Delta\pm\Delta)t}\over {(\omega-\Delta\pm\Delta)^2}}\nonumber\\
\eta_{\pm, k_*}^2(1-{\sin(\omega r/c)\over (\omega r/c)})
(\coth{\omega\over 2T}-\tanh{\omega-\Delta\over 2T})
\eea
where $k_*\sim 1/r$. It is also worth noting that for a broadly distributed entanglement 
($\rho_{JW}(x,0)\approx const$) the decoherence rate 
can be more difficult to deduce.

Representing the total decoherence rate (14) as a sum $\Gamma=\cos^22\theta_{k_*}
\Gamma_++\sin^22\theta_{k_*}\Gamma_-$, we separate between 
the processes of scattering with almost no energy exchange 
($\omega\approx 0$), thus regarding $\Gamma_+$ as a loose analog of the "pure dephasing" 
rate in the conventional single-qubit problem, and finite ($\omega\approx 2\Delta$) 
energy transfers
characterized by the "relaxation" rate $\Gamma_-$.  

Consistent with this interpretation, $\Gamma_-$ also appears to control
the (vanishing with decreasing $J$ and/or increasing $\Delta$, see below) 
decay rate of the diagonal matrix elements of ${\hat \rho}_{ij}(t)$,
as manifested by the correlator $<S^{z}_i(t)S^{z}_j(0)>$
related to the statistically
averaged JW fermion amplitude
$<\chi^\dagger_i(t)\chi_i(t)\chi^\dagger_j(0)\chi_j(0)>$.

Performing the integration in Eq.(14), we obtain  
$\Gamma_+(r,t)\propto T(min[r/c,t])^{2-\alpha}$ at long times/distances
($min[r/c,t]>1/T$). In the opposite limit ($max[r/c,t]<1/T$)  
$\Gamma_+(r,t)$ behaves as $\propto T^{3+\alpha}r^2t^2$, while
at intermediate scales ($min[r/c,t]<1/T<max[r/c,t]$)
 $\Gamma_+(r,t)\propto T^{1+\alpha}min[r/c,t]^2$.

In the critical ($\Delta=0$) regime $\Gamma_-(r,t)$, too, is
given by the above asymptotics, although a finite gap limits their applicability
to $max[r/c,t]<1/\Delta$, while at $min[r/c,t]>1/\Delta$ Eq.(14) yields
$\Gamma_-(r,t)\propto Tt\Delta^{\alpha-1}$.

Thus, at $\alpha=1$ corresponding to the Ohmic regime
in the conventional spin-bath model the total 
decoherence rate $\Gamma$ turns out to be essentially independent of $\Delta$, as long as 
$\Delta\lesssim T$, thus implying
that at sufficiently high temperatures 
the quality of pairwise entanglement can be largely insensitive to 
the proximity to the critical point 
(the speed $v$ of the ballistic entanglement spreading  
does decrease with increasing $\Delta$, though).
However, in the case of a
sub-Ohmic ($0<\alpha<1$) environment 
the decoherence rate (14) as a function of the gap develops a minimum at 
\be
\Delta\sim(1-\alpha)^{1/1-\alpha}t^{-1}
\ee
Therefore, for a given time interval $t$ between consecutive gate operations 
one can reduce the environmentally induced
decoherence by tuning the system farther away from criticality
and opening up the spectral gap (15). 

It also follows from Eq.(14) that, rather naturally, the decoherence
rate decreases at smaller spatial separations between the qubits (large $k_*$),
thus further emphasizing the special importance of the 
(next-) nearest-neighbor entanglement \cite{Amico}.

In summary, we studied the quality of time evolution between 
gate operations ("quantum memory") in a $1D$ array of $N\gg 1$ coupled 
qubits which are also subject to a spatially non-iniform (sub-)Ohmic dissipative environment.
We found that the shared quantum entanglement
that reaches its maximal
attainable
value at the critical points 
of the qubit-chain Hamiltonian \cite{Amico} is generally accompanied
by a concomitant increase in the entanglement between the qubits and the environment.
Therefore, the requirement of preserving the qubits' entanglement over a certain idling time 
between consecutive gates can be better 
fulfilled away from criticality and/or at smaller separations between the 
pair of qubits constituting a gate.

Viewed in the context of their potential applications to 
quantum information processing, our results suggest that, in contrast to the 
expectation expressed in the first of Refs.\cite{Amico}, the mere onset of
the long-range correlations in the near-critical regime  
does not guarantee that an interacting multi-qubit system 
would autmatically become more robust against noise. 
In fact, the inter-qubit couplings appear to control both the onset of a critical behavior
and the dependence of the decoherence rates on the proximity to a critical point.
Therefore, when choosing the parameters of the qubit Hamiltonian (1),
one will have to optimize between the rather 
contradictory criteria in order for a qubit-chain prototype of a quantum register 
to achieve its target performance in terms of both, the high degree
of entanglement and reliable coherence control. 

This research was supported by ARDA under Contract DAAD19-02-1-0049
and, in part, by NSF under Grant DMR-0071362.

\wt

\begin{references}


\bibitem{Zanardi} P. Zanardi and M. Rasetti, Phys. Rev. Lett.
{\bf 79}, 3306 (1997).

\bibitem{Viola} L. Viola, E. Knill and S. Lloyd, Phys. Rev. Lett. {\bf 82}, 2417 (1999).
 
\bibitem{Amico} A. Osterloh, L. Amico, G. Falci and R. Fazio, Nature {\bf 416}, 608 (2002);
T. J. Osborne and M. A. Nielsen, Phys. Rev. {\bf A66}, 032110 (2002).

\bibitem{Wooters} S. Hill and W. K. Wooters, Phys. Rev. Lett. {\bf 78}, 5022 (1997).

\bibitem{Barouch} B. M. McCoy, E. Barouch and D. B. Abraham, 
Phys. Rev. {\bf A4}, 2331 (1971).

\bibitem{DVK} D. V. Khveshchenko,
Phys. Rev. {\bf B65}, 235111 (2002); Nucl. Phys. {\bf B642}, 515 (2002).

\bibitem{Eberle} J. S. Pratt and J. H. Eberle, Phys. Rev. {\bf A64}, 195314 (2001).

\bibitem{Reina} W. G. Unruh, Phys. Rev. {\bf A51}, 992 (1995);
G. M. Palma, K. A. Suominen and A.K. Ekert,
Proc. R. Soc. London, Ser.{\bf A452}, 567 (1996);
L. M. Duan and G. C. Guo, Phys. Rev. {\bf A57}, 737 (1998);
J. H. Reina, L. Quiroga and N.F. Johnson, Phys. Rev. {\bf A65}, 032326 (2002).


\end{references}
\end{document}